\pdfoutput=1
\documentclass[pra,twocolumn,showpacs,superscriptaddress]{revtex4}
\usepackage{graphicx}
\usepackage{amsmath}
\newcommand{\n}{\nonumber}
\newcommand{\bn}{\begin{eqnarray}}
\newcommand{\en}{\end{eqnarray}}
\newcommand{\eml}{\end{multline}}
\newcommand{\bml}{\begin{multline}}
\newcommand{\h}{\hspace}

\begin{document}

\title {Quantum Paddlewheel with Ultracold Atoms in Waveguides}
\author{Kunal. K. Das}
\affiliation{Department of Physical Sciences, Kutztown University of Pennsylvania, Kutztown, Pennsylvania 19530, USA}
\affiliation{Kavli Institute for Theoretical Physics, UCSB, Santa Barbara, CA 93106}
\author{Matthew R. Meehan }
\affiliation{Department of Physical Sciences, Kutztown University of Pennsylvania, Kutztown, Pennsylvania 19530, USA}
\author{Andrew J. Pyle}
\affiliation{Department of Physical Sciences, Kutztown University of Pennsylvania, Kutztown, Pennsylvania 19530, USA}

\date{\today }
\begin{abstract}
We propose and study a quantum pump which emulates a traditional paddlewheel, that can be implemented with ultracold atoms in waveguides. We use wavepacket propagation to study its single-mode dynamics, which also determines its multimode current for mesoscopic setups. Energy flow with or without particle transport is possible. The spectrum reveals unusual features such as nonuniform Floquet side-bands and counter-intuitive scattering. Explanations are found by examining the scattering dynamics comparatively using \emph{quantum, classical} and \emph{semiclassical} pictures, indicating a rich system and experimentally accessible method to explore quantum versus classical dynamics.
\end{abstract}
 \maketitle

\date{\today }

\section{Introduction}
The paddlewheel is among the oldest human inventions for harnessing natural flow for useful work. We study a quantum-mechanical version of a paddlewheel that can be similarly useful in generating and controlling coherent currents. A specific motivation is the long-standing interest in the phenomenon of quantum pumps \cite{thouless} that has eluded clear implementation in mesoscopic electronics. The general motivation is to probe the scattering dynamics of clouds of ultracold atoms colliding with localized external potentials that vary in time.

Quantum pumps originated as a means of generating quantized transport of charge using a topological invariant \cite{thouless, Niu-Thouless}, but advances in nanotechnology in the decades since have shifted the interest towards generating controlled unidirectional flow of charge \cite{Kouwenhoven-PRL-1991,Aleiner-Andreev-1998,brouwer-1,buttiker-floquet,turnstile,das-opatrny,Zhou-Spivak-Altshuler,kim-floquet,
avron-geometry-PRB,nanostructures,arrachea-green,Aono, zhou-mckenzie,Makhlin-Mirlin-2001,entin-wohlman-SAW}, spin \cite{chamon-spin,watson,blaauboer} and even entanglement \cite{das-PRL,samuelsson-buttiker} in mesoscopic circuits by means of time-varying potentials. Theoretical interest in the subject has continued to be strong, expanding its scope to include superconductors \cite{Giazotto-superconductor} and graphene \cite{Blaauboer-graphene} and carbon nanotubes \cite{Buitelaar-nanotube,buitelaar-nanotube-2} in recent years. This sustained interest is particularly remarkable because experimental demonstration of adiabatic charge pumps in normal mesoscopic conductors has stubbornly remained elusive \cite{switkes,brouwer-2}, although there has been varied success involving spin currents\cite{watson}, hybrid normal-superconducting systems \cite{Giazotto-superconductor} and carbon nanotubes \cite{buitelaar-nanotube-2}.

Despite their wide range and number, studies of quantum pumps have always been in the context of electronic systems, until recently when one of us proposed a way to implement them with trapped ultracold fermionic atoms \cite{Das-Aubin-PRL2009}; neutral carriers bypass the primary complications facing electron pumps \cite{brouwer-2}. In a subsequent study \cite{Das-wavepacket}, we developed a novel approach for simulating general mesoscopic transport (including quantum pumps) with wavepackets of ultracold bosons, which can access details of transport dynamics at a single mode level instead of the usual multimode average inherent for electrons.

In this paper, we study a paddlewheel pump with our wavepacket approach, but broadening its scope significantly to simulate the scattering dynamics classically as well as quantum mechanically in an equivalent treatment that allows direct comparison, for which we use a semiclassical picture. This enables a direct visualization of the dynamics at play in quantum pumps; yet allows easy simulation of the multi-mode fermionic current relevant for an electronic paddlewheel. Semiclassical methods are not new to the study of electronic quantum pumps, notably used to study a classically chaotic pumps \cite{Brouwer-semiclassical} where current is generated entirely by quantum interference, and to examine the limit of a large number of transport channels \cite{Mucciolo-semiclassical}. Our semiclassical treatment is differently motivated, being concerned with wavepackets of ultracold atoms in single-channel operation without chaos, with the main purpose of distinguishing classical features from purely quantum features in quantum pumps. This broadens the scope of quantum pumping from an esoteric transport mechanism in nano-electronics to be a promising test-bed for examining quantum-classical paradigms with ultracold atoms. A paddlewheel pump is particularly well-suited for that purpose since it generates significant current in \emph{both} classical and quantum operations unlike the popular turnstile pump \cite{buttiker-floquet, turnstile,das-opatrny}.

The paddlewheel mechanism considered here is distinct from typical quantum pumping mechanisms, which involve at least two spatially separated independent potentials like in a turnstile pump \cite{Kouwenhoven-PRL-1991,Aleiner-Andreev-1998, brouwer-1,buttiker-floquet,turnstile} or  use traveling potential waves \cite{entin-wohlman-SAW, buitelaar-nanotube-2}.  The paddlewheel requires only one potential barrier, making it perhaps the simplest of quantum pumps with two time-varying parameters, easier to implement and study in atomic systems. Time-varying single-barrier potentials have been considered previously for electron transport, but in closed ring geometry \cite{Sela-Cohen-PRB-2008,Rosenberg-Cohen-JPA} where the operation is  dubbed `quantum stirring' to underscore significant differences from quantum pumps that operate in open systems. Such single barrier stirring mimics a piston \cite{Chuchem-Dittrich-Cohen} with a lateral back and forth motion, impossible in a real paddlewheel, which intrinsically has a discontinuous reset, as each new paddle enters the medium at the same position. The discontinuity has crucial relevance for several distinct features of a paddlewheel, such as counterintuitive scattering.

We describe our physical model in Sec. II, and study the features of single-mode and multimode current generated by a paddlewheel in Sec. III. Their prominent features are then examined by analyzing the quantum scattering spectra and momentum distributions in Sec. IV and compared with classical and semiclassical distributions in Sec. V. Different counter-intuitive scattering features are described and explained in Sec VI, and we conclude in Sec. VII with a demonstration of experimental feasibility and a discussion placing our results in the context of past studies and future possibilities.

\begin{figure}[t]
\includegraphics[width=0.9\columnwidth, angle=-90]{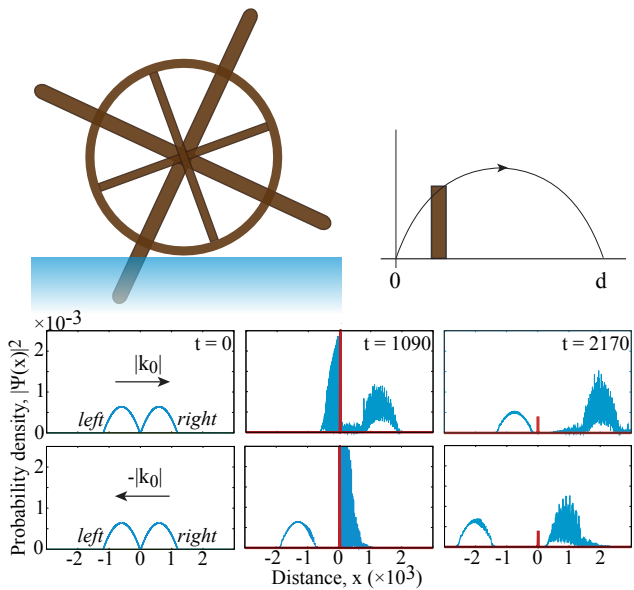}
\caption{Schematic of a paddlewheel and its implementation with a translating potential barrier with oscillating strength. Panels show snapshots of our simulation with wavepackets (starting on left and right of the barrier) with positive  (upper) and negative (lower) incident momenta $p_0=\hbar k_0$. In each cycle, the barrier starts from zero amplitude at the origin and translates a span $d$ to the right and resets(the distance $d$ is imperceptible on the scale of these snapshots).  Approximate snapshot times are shown and parameters and units used are defined after  Eqs.~\ref{potential} and \ref{wavepacket}.}\label{schematic}
\end{figure}

\section{Physical Model}

A mesoscopic quantum pump uses time-varying potentials to generate directed flow through quasi one-dimensional (1D) nano-wires connected \emph{without} bias to macroscopic contacts that act as source and absorbing reservoirs for fermionic carriers.  Ballistic motion \cite{Ferry-Goodnick} is assumed in the wires, so the current is determined by the scattering at the potential. The net current can be defined as the integral over contributions at each incident momentum $p_0=\hbar k_0$,
\bn\label{fermion-current} J_F(t)&=&\int_{-\infty}^{\infty}\frac{dk_0}{2\pi}f(k_0) J(k_0,t)\\ J(k_0,t)&=&(\hbar/m) \langle
\psi(k,t)|k|\psi(k,t)\rangle/\langle \psi(k,t)|\psi(k,t)\rangle,\n\en
$f(k_0)$ is the Fermi distribution function and $\psi(k,t)$ the scattering wavefunction generated by incident $e^{ik_0x}$. The single mode current can therefore be defined to be
\bn J_s(k_0,t)=\frac{1}{2}[J(+|k_0|,t)+J(-|k_0|,t)].\en The incoherent sum captures the lack of coherence between particles from different reservoirs. We simulate this by representing carriers at each incident mode with wavepackets with the appropriate velocity that interact with the potential to yield the scattered wavefunction. Broad wavepackets, that interact with the time-varying potential for several cycles, emulate the usual assumption of plane waves in standard approaches. The multimode current of fermions can be computed by sampling the single-mode current over the relevant range of incident momenta and approximating the integral in Eq.~(\ref{fermion-current}) by a Riemann sum.

This model can be directly implemented in systems of ultracold atoms in quasi-1D waveguides. Start with a wavepacket $\psi(x,t=0^-)$ of ultracold atoms in an axial trap \cite{Das-Aubin-PRL2009}, centered about (or close to) the scattering potential created by a tightly focused blue-detuned laser. Transport is initiated by turning off the axial trap and giving the atoms the appropriate momentum $\pm \hbar k_0$ using Bragg beams \cite{Ketterle}, $\psi(x,0^+)=e^{\pm ik_0x}\psi(x,0^-)$.
The wavepacket is allowed to evolve until the scattered wavepacket has cleared the potential barrier, and then the spatial and momentum distribution are imaged.

In a classical paddlewheel shown in Fig.~\ref{schematic}, the vertical extent of a paddle inside water is $L\sin(\omega t)-L_0>0$ where $L$ is the distance of the paddle tip to the center located $L_0$ above the surface; $\omega$ is the angular velocity. We capture the essential features with a time-varying potential
\bn V(x,t)=U_0e^{-(x-f(t))^2/(2\sigma^2)}[1+\sin(\omega_{osc} t +\phi)].\label{potential}\en
Like typical quantum pumps, it has two independent time-varying parameters - the barrier height oscillating with frequency $\omega_{osc}$, and the barrier position, $f(t)=\mod(vt,d)$ resetting after distance $d=v\times2\pi/\omega_{tran}$. We set $\phi=3\pi/2$ and $\omega_{osc}=\eta\times\omega_{tran}$ with integer $\eta$, so for $\eta=1$ the barrier tip traces a curve shown schematically in Fig.~\ref{schematic}: As the barrier vanishes at $x=d$ at the end of a cycle it re-emerges at $x=0$ like the next paddle entering the medium ($\eta>1$ emulates a $\eta$-wheel chain, one paddle  each, entering sequentially). The Gaussian barrier is used in most of our simulations since it models a typical laser profile, but in some specific cases we will find it useful to also use a rectangular barrier to help with our analysis.

\emph{Our results are dimensionless}, set by choice of energy ($\epsilon$), length ($l$) and time ($\tau$). For comparison with physical parameters, we choose the transverse harmonic trap frequency of the waveguide $\omega_r$ to define our units  $\epsilon=\hbar\omega_r$, $l=\sqrt{\hbar/m\omega_r}$ and $\tau=\omega_r^{-1}$, with $m$ being the mass of individual atoms. The transverse frequency is convenient as a parameter that does not change during the evolution, and also this choice of units yields a form of the Schr\"{o}dinger equation that is equivalent to setting $\hbar=m=1$. The same numerical values of the units are assumed for our classical simulations as well, for consistent comparison. Unless otherwise mentioned, in most of our simulations we use barrier parameters $U_0=0.5,v=1.5,\sigma=5,\omega_{osc}=0.2$, which set the reset span to be $d=\eta\times 47.1$.  Our main considerations in picking them was to yield realistic experimental parameters that we discuss later, and to prominently display the features of interest.

\section{Current and Particle Transport}

\begin{figure}\includegraphics[width=\columnwidth]{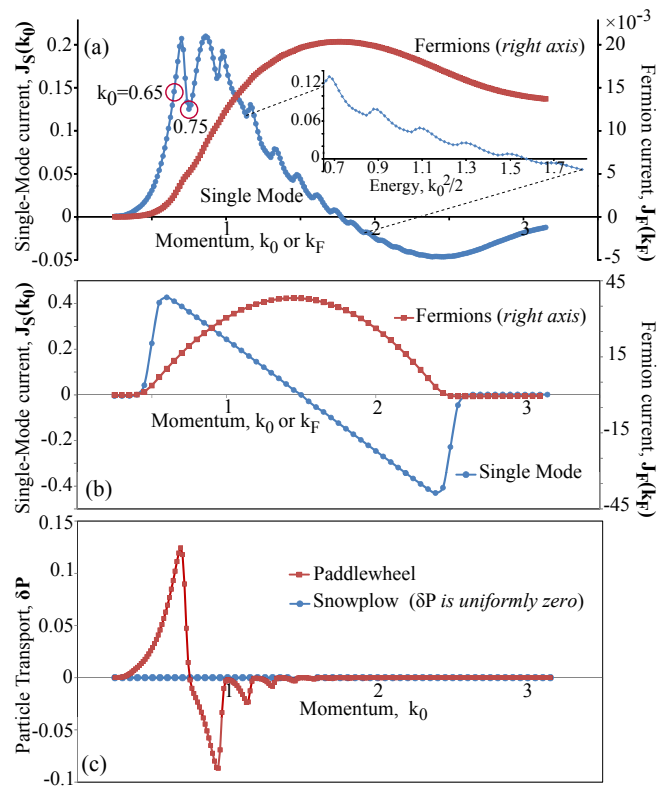}
\caption{(a) Single-mode current versus incident momentum ($k_0$) and multi-mode (fermionic) current versus Fermi momentum ($k_F$) for a paddlewheel pump. Wavepackets widths $b=400,600,1000$ yield the same plot. The prominent dip is washed out in the fermion average. [\emph{Inset}] On plotting the current versus incident energy, the oscillations display the periodicity $\omega=0.2$ of the barrier resetting.  (b) For comparison the current is plotted for exactly the same parameters for a snowplow (barrier translating without resetting or oscillation). (c) Plot of net particle transport $\delta P$ for paddlewheel and snowplow. }\label{current}
\end{figure}

Primary dynamical information about a paddlewheel pump can be surmised from the current it generates.  In Fig.~\ref{current}(a), we plot $J_s(k_0)$ as a function of the incident momentum, as well as the integrated fermionic current $J_F(k_F)$ as a function of the Fermi momentum ($f(k_0)=[1-\Theta(k_F)]\Theta (-k_F)$, assuming complete degeneracy). We use initial Thomas-Fermi envelopes \bn \rho_0(x)=|\psi(x,t=0^-)|^2=[b^2-(x-x_m)^2],\label{wavepacket}\en
with packet radius $b\geq|x-x_m|$; since small interactions can broaden degenerate bosonic clouds into Thomas-Fermi profile without impacting the linearity of the dynamics \cite{Das-wavepacket}. However, the precise shape is unimportant as long as the packets are much wider than the barriers and allows interaction over several cycles: The plot in Fig.~\ref{current}(a) overlays simulations with three different widths of the incident packets (b=400,600,1000), and the current profiles are indistinguishable; implying equivalence to using plane-waves. Therefore, in all our other simulations, we only use b=600. In the quantum simulation, the wavepackets are propagated by the Schr\"{o}dinger equation using a split-step operator method \cite{Agarwal}. This assumes that interactions among atoms are made sufficiently small by using dilute gas and by such means as Feshbach resonance \cite{Grimm-RMP-Feshbach} to be able to neglect non-linear effects.

Current is generated due to (i) redistribution of momenta and/or (ii) net particle transport gauged by %
\bn\delta P=\int_0^\infty dk |\psi(k)|^2-\int_{-\infty}^0 dk |\psi(k)|^2.\en
Comparison of the current profile of the paddlewheel with Fig.~\ref{current}(b) for a `snowplow' pump \cite{avron-math,das-opatrny}, a barrier translating  uniformly without oscillating or resetting, highlights the skewness of the paddlewheel current towards lower incident momenta.  Momentum redistribution occurs for both: For the snowplow, Galilean transformation shows that incident momenta $\pm |k_0|$ lead to four  scattering peaks at $\pm |k_0|$ and $\mp |k_0|+2v$. However, plots of the net particle transport in Fig.~\ref{current}(c) shows that $\delta P=0$ for a snowplow but \emph{not} for the paddlewheel particularly at low incident momenta. The implication is that paddlewheel pump can be used to \emph{selectively pump energy with or without net particle transport}. We found this is true even when when the oscillation of the barrier is turned off, $\omega_{osc}=0$, indicating that the particle transport arises from multiple interactions with the barrier due to it resetting; this is more prone to occur at lower incident momenta $k_0$, due to longer dwell time in the region of the potential.

\begin{figure}[t]\includegraphics[width=0.83\columnwidth,angle=-90]{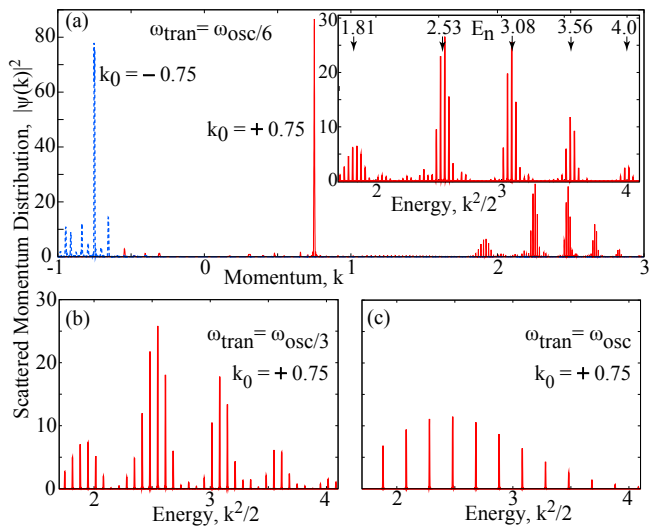}
\caption{ (a) Quantum momentum distribution \cite{movies} for a paddlewheel with $\omega_{tran}=\omega_{osc}/6$ shows differential scattering for $\pm |k_0|$. [\emph{Inset}] Partial energy spectrum: The fine combing due to resetting has uniform spacing of $\omega_{tran}$, but the five clusters due to oscillation \emph{are not} equally spaced by $\omega_{osc}$, but are instead centered about  $E_n,n=0,\pm1,2,3$ (shown) as predicted by Eq.~(\ref{Floquet}). (b) The clusters (and the combing) broaden as reset frequency is increased to $\omega_{tran}=\omega_{osc}/3$ and (c) merge when the frequencies are equal $\omega_{tran}=\omega_{osc}$ and all the peaks become equally spaced.}\label{momentum-dist}
\end{figure}

The single-mode current profile in Fig,~\ref{current}(a) displays oscillations, caused by the barrier resetting: since we find them to persist even with barrier oscillations turned off. Moreover, when the current is plotted [\emph{inset}] as a function of the incident energy $k_0^2/2$, the oscillations have the same periodicity $\omega_{trans}=0.2$ as for the barrier reset.

But, the most prominent feature of Fig.~\ref{current}(a) is the significant dip in the single-mode current that occurs at incident momentum of $k_0=0.75$ as marked in the figure. It occurs precisely where the current profile would have had its global peak following its pattern of oscillations. It is highly relevant that all these features are conspicuously washed out in the multimode current. This indicates underlying physics that cannot be accessed in multimode electronic quantum pumps, and will show the advantage of wavepackets in providing physical explanations.

\section{Quantum distribution}
The scattered wavepackets in momentum space give the momentum distribution and hence the spectrum. Contributions due to $+|k_0|$ and $-|k_0|$ incident momenta can be distinguished, as shown for a single-mode paddlewheel pump in Fig.~\ref{momentum-dist}(a), exposing the differential scattering that leads to unidirectional net flow. The spectrum displays the \emph{intertwined} effects of the different motions of the potential - \emph{translation, oscillation} and \emph{resetting}. The distribution is clustered about the scattered momenta due to translation,  $\pm |k_0|$ and $-|k_0|+2v$, that would occur for a snowplow pump as follows from a Galilean transformation \cite{das-opatrny}; a possible fourth cluster at $|k_0|+2v$ is absent since there is negligible reflection of particles incident with negative momenta for the values of the parameters used.  The oscillations ($\omega_{osc}$) create sub-clusters (5 are seen clearly about $-|k_0|+2v=2.25$) and finally the resetting ($\omega_{tran}$) creates fine combing within each such sub-cluster.

\begin{figure}[t]\includegraphics[width=0.8\columnwidth,angle=-90]{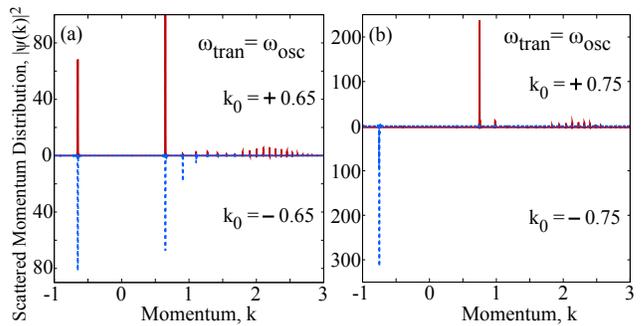}\vspace{-2.5cm}
\caption{ Quantum scattered momentum distribution corresponding to incident momenta $k_0$ around the dip in Fig.~\ref{current}(a) (circled in that figure), for (a) $k_0=\pm 0.65$ and (b) $k_0=\pm 0.75$. For clarity, scattered distribution due to incidence from left (right) with $+|k_0|(-|k_0|)$ are shown above (below) axis. Note the presence in (a) and absence in (b) of the reflection peaks at $\mp |k_0|$ due to incident $\pm |k_0|$.}\label{dip-dist}
\end{figure}

The scattering energy spectrum in Fig.~\ref{momentum-dist}(a)[inset] seems to partially violate Floquet's theorem of \emph{uniform} spacing of side-bands about the incident energy $E_0=k_0^2/2$. While the fine-combing, due to the resetting, is indeed uniformly spaced by $\omega_{tran}$; the  subclusters, due to the oscillation, are \emph{not equally spaced} by $\omega_{osc}$. We found that the correct way to understand the sub-clusters is to consider them in a frame \emph{translating with the barrier}, where the incident particles will have energy $\bar{E}_0=(k_0-v)^2/2$. The subclusters have the expected uniform Floquet spacing of $\omega_{osc}$ about this energy value. Therefore, in the rest frame, the sidebands due to the oscillation occur at
\bn k_n&=&v\pm sign(k_0-v)\sqrt{(k_0-v)^2+2n\omega_{osc}}\ ,\label{Floquet}\en
causing non-uniform spacing of $E_n=k_n^2/2$ due to the cross-term.  The $\pm$ are for transmission/reflection.  As indicated in Fig.~\ref{momentum-dist}(a)[inset], the values of $E_n$ computed for $n=0,\pm1,0,2,3$ exactly match the observed locations of the sidebands.

As the reset period is increased to $\omega_{tran}=\omega_{osc}/3$ [Fig.~\ref{momentum-dist}(b)], the fine combing peaks spread out and the clusters due to the oscillation $\omega_{osc}$ broaden, so when $\omega_{tran}=\omega_{osc}$ [Fig.~\ref{momentum-dist}(c)], all the peaks become uniformly spaced, and the effects of oscillation and resetting in the spectrum become indistinguishable. This last case coincides with most of our simulations since we use $\eta=1$.

\begin{figure}\includegraphics[width=\columnwidth]{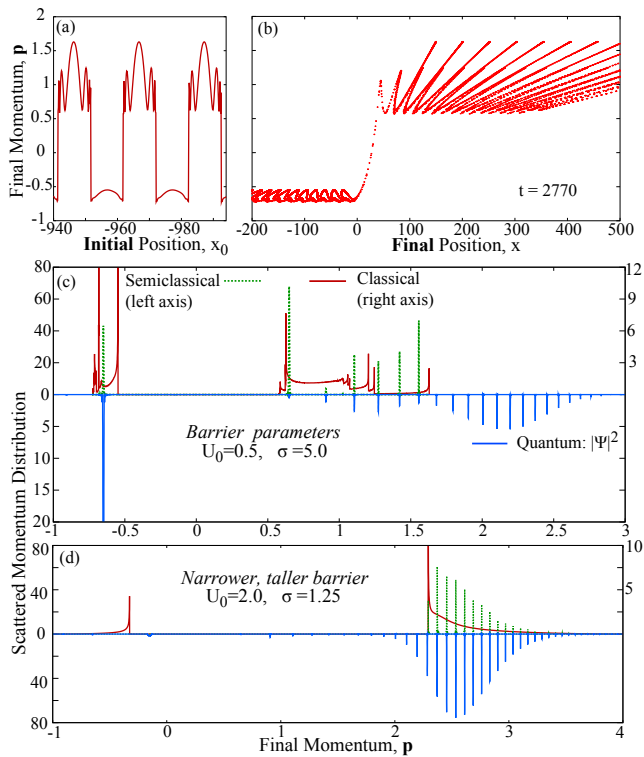}\vspace{-1cm}
\caption{ Classical simulation \cite{movies} of the paddlewheel: Final momentum $p$ versus (a) \emph{initial} position $x_0$ and (b) \emph{final} position $x$; due to particles starting on left of the barrier with $k_0=0.65$.  (c) Scattered momenta distributions with classical and semiclassical above axis and quantum below axis. (d) Using narrower (but taller) barrier makes the peak at $-k_0$ essentially disappear.}\label{analysis}
\end{figure}

We now seek the cause of the sharp dip [Fig.~\ref{current}(a)] in $J_s$ by plotting the scattered momentum distribution in its neighborhood in Fig.~\ref{dip-dist} (distributions due to left and right incidence are plotted above and below axes to easily differentiate them). We see something surprising: The distribution Fig.~\ref{dip-dist}(b) corresponding to the bottom of the dip $|k_0|=0.75$ agrees with our discussion above. However, the distribution in Fig.~\ref{dip-dist}(a) for $|k_0|=0.65$, just below where the dip occurs, for both $\pm|k_0|$ incident momenta, we see additional prominent peaks indicating \emph{reflection} at the \emph{same magnitude} as the incident momenta ($\mp |k_0|$ for $\pm |k_0|$). We found this to be true for lower values, $|k_0|<0.65$ as well, but missing for momenta above that at the dip. This is puzzling because the primary reflection peaks should be only at $\mp |k_0|+2v$.  Is this a purely quantum effect, or does it have classical roots? We examine that in the next section.

\section{Semiclassical and Classical distribution}

We simulate the same scenario classically with ensembles of particles with initial position and spatial distribution matching  the quantum wavepacket, and having constant initial momentum $k_0$ (the momentum spread of the spatially broad quantum packets used here is very small and has negligible effects). In order to unambiguously identify the scattered peaks, we use only one incident packet [refer to Fig.~\ref{schematic}] for each momentum - the packet approaching the barrier (starting on the left of it for $+|k_0|$ and starting on the right for $-|k_0|$). The receding wavepackets have far less interaction with the barrier in any case. We do the same for the corresponding quantum simulations in Figs.~5,6 and 7, for consistent comparison.   For the classical propagation, we compute the trajectories by numerically integrating the classical equations of motion:
\bn \frac{dx}{dt}=\frac{\partial H}{\partial p};\h{2mm}  \frac{dp}{dt}=-\frac{\partial H}{\partial x};\h{2mm} \frac{d\tilde{S}}{dt}=-x\frac{d p}{d t}-H,\en
$H$ is the Hamiltonian and $\tilde{S}(p,t)$ is the momentum-space equivalent of the classical action. The semi-classical wavefunction at a time $t$ can then be constructed
\begin{equation}
\psi_{SC}(p,t)={\textstyle \sum_j}\rho_0(x^j_0)\left| \partial p/\partial x_0\right|_{x^j_0} ^{-\frac{1}{2}}e^{i[\tilde{S}_j(p,t)/\hbar-\mu_j\pi/2]}\label{semi-classical}
\end{equation}
where $\mu_j$ is the Maslov index \cite{Maslov,Das-single-barrier}. Since the potential is periodic, the final momentum $p$ is a periodic function of the initial position, with each $p$ having contributions from multiple initial positions $x_0(p,t)$, as evident in Fig~\ref{analysis}(a), where a horizonal slice corresponding to a given momentum intersects the curve at multiple points.  It is clear from that figure, that even within each period there are multiple branches arising from particles incident on the barrier at different segments of a cycle that still lead to the same final momentum. Therefore, the sum in Eq.~(\ref{semi-classical}) is over both (i) intra-cycle branches within a cycle and (ii) inter-cycle repetition due to the periodicity. In that equation for the semi-classical wavefunction, the initial density determines $\rho_0$, the local slope in Fig.~\ref{analysis}(a) of $p$ versus $x_0$ determines the Jacobian $\left| \partial p/\partial x_0\right|$; and  Fig.~\ref{analysis}(b) of $p$ versus \emph{final} position $x$  fixes the Maslov index as follows: Setting $\mu_j=1$ at the first momentum branch, at each turning point where $dp/dx=0$, it is incremented by $+1$ or $-1$ for clockwise or counterclockwise turns respectively.

The \emph{semiclassical} momentum distribution is given by $|\psi_{SC}(p,t)|^2$, which includes interference among the different classical trajectories. If the terms in Eq.~(\ref{semi-classical}) are absolute-squared \emph{before} summing, the phase information is lost and we obtain the \emph{classical} distribution. Both are compared with the quantum distribution in Fig.~\ref{analysis}(c).  The semiclassical picture is useful to relate the classical with  the quantum features. As defined, the semiclassical distribution is confined to the classically allowed regime, where it displays periodic peaks that \emph{coincide in position} with the Floquet peaks. The peak heights are not in general agreement, since features like quantum tunneling are not included in our semiclassical picture. Notably, the Maslov index affects peak heights but not their locatio; we demonstrate that by approximating it by using $(x_0,p)$ chart in Fig.~\ref{analysis}(c), but then we use the $(x,p)$ chart for Fig.~\ref{analysis}(d)to get better agreement with quantum peak heights. In both cases, there are some quantum peaks beyond the classically allowed momenta, which cannot be captured by our semiclassical function since in those regimes its amplitude is identically zero. Additional features \cite{Das-single-barrier} need to be added to the semi-classical picture to incrementally describe those, which is not essential for the purpose of this paper.  The primary conclusion here is that Fig.~\ref{analysis}(c) clearly indicates that the anomalous reflection by a paddlewheel at $- k_0$ due to particles incident with monenta $k_0$  \emph{occurs classically too}, with a well-defined semiclassical peak exactly where we see the quantum peak.  Furthermore, we can also conclude that this is an effect of the finite width of the barrier, since making the barrier narrower and proportionately taller makes the anomalous peaks disappear as we see in Fig.~\ref{analysis}(d).

\begin{figure}[t]\includegraphics[width=\columnwidth]{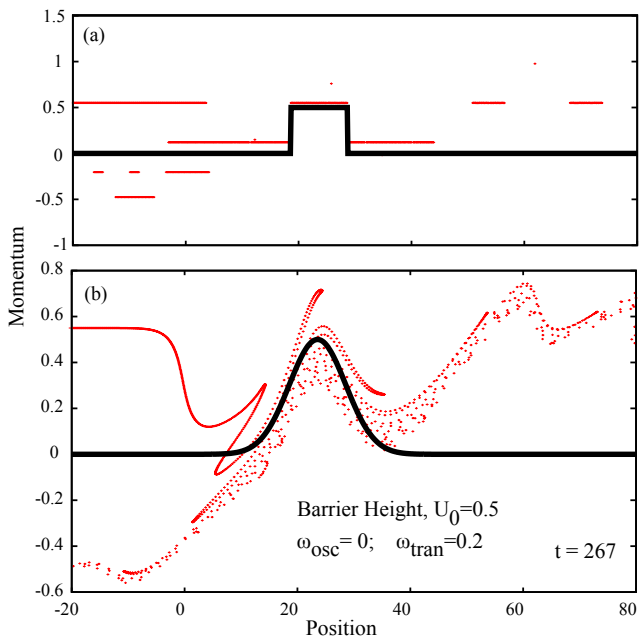}\vspace{-2.5cm}
\caption{ Snapshots \cite{movies} of a classical stream of particles incident from \emph{left} at $k_0=0.55$ on a modified paddlewheel that resets but does not oscillate, $\omega_{osc}=0$ (and also $\phi=0$ to have a non-vanishing barrier): (a) rectangular barrier and (b) Gaussian barrier. Both show the classical origins of the anomalous peak at $-k_0$ in Fig.~\ref{dip-dist}(a). The parameters shown are common to both figures; for the rectangular barrier the width is $2\sigma=10$.}\label{classical}
\end{figure}

\section{Counter-intuitive scattering}

Having established that the anomalous reflection peaks have classical roots, we examine the time-evolution of the classical ensemble, using snapshots of the scattering such as shown in Fig.~\ref{classical}.  We found that the anomalous peaks appear even with no oscillation, hence we simplify the scenario, by turning it off by setting $\omega_{osc}=0$ in Eq.~(\ref{potential}), and also $\phi=0$,  to keep the barrier at a constant height as it translates and resets. In this case, the dynamics is particularly transparent when the barrier is rectangular as shown in Fig.~\ref{classical}(a); specifically we see discrete strips of particles being reflected. These particles were elevated by the barrier as it resets, gaining potential energy that transforms back to kinetic energy as they roll off the barrier. The resulting momenta as they roll off are given by
\bn k=v+ sign(k_0-v)\sqrt{(k_0-v)^2+2nU_0}\en
where $n$ is the number of times a particle interacts with the barrier, each time gaining energy equivalent to its strength $U_0$. Notably, when $0<k_0<v$, the particles roll off the trailing edge of the barrier and get a kick in the \emph{negative} direction so that the final momentum is reduced $k<k_0$. So \emph{counter-intuitively, the momentum decreases although the barrier adds energy}, and with multiple interactions, momentum can be reversed. Remarkably, the semi-classical picture shows that once negative momenta are significantly populated, interference among classical trajectories create a peak at $-k_0$ where we find the anomalous Floquet peak in the quantum scattering distribution. Similar arguments apply when $\omega_{osc}\neq 0$, $k_0<0$, and with Gaussian paddlewheels seen in Fig.~\ref{classical}(b), where the main difference is that the scattering pattern is continuous instead of being in discrete strips.

This also explains why the anomalous peak disappears at higher incident momenta: If $k_0$ is sufficiently high there are not enough interactions for the momenta to change sign, and such a peak does not emerge, thus our anomalous peaks disappear above $k_0\simeq 0.70$, causing the dip in Fig.~\ref{current}. It is also clear that a finite width of the barrier is essential, otherwise the potential energy gain by the `elevator' effect would not impact significant number of particles when the barrier resets.

\begin{figure}[t]
\includegraphics[width=\columnwidth]{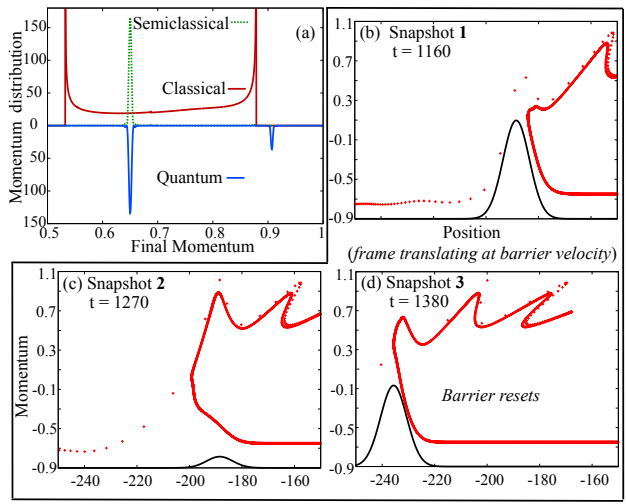}
\caption{(a) Counter-intuitive scattering: The scattered momentum distribution for particles incident on a paddlewheel from the \emph{right} at $k_0=-0.65$, with classical and semiclassical shown above axis and quantum below. Although the maximum potential is less than the relative incident energy, $2U_0< (k_0+v)^2/2$, there is almost total reflection. (b-d) Snapshots of \emph{classical} scattering show that a \emph{Sisyphus} effect, due to multiple interactions with the barrier, causes the reflection \cite{movies}.}\label{Fig-7}
\end{figure}

The paddlewheel pump also demonstrates another distinct and different anomalous scattering phenomena that is illustrated in Fig.~\ref{Fig-7}: Particles are incident on the barrier from the right with negative incident momentum $-|k_0|$, and in panel (a) we see that there is almost \emph{total reflection} in both classical and quantum scenarios. This is remarkable because the incident kinetic energy of the particles significantly exceeds the maximum height of the paddlewheel barrier, $|k_0+v|^2/2>2U_0$, for the parameters used. So classically there should be complete transmission, and even quantum mechanically transmission should be almost complete, yet we see quite the opposite.
Unlike the previously described effect that was due to resetting, here both oscillation and resetting play a role. This effect can also be understood using classical physics: The particles loose kinetic energy as they climb up the barrier; and because the barrier resets discontinuously, they may not get a chance to gain it back by rolling down the other side. A repetition of this, due to multiple interactions with the barrier then leads to a \emph{Sisyphus} effect, where the particles climb the barrier repeatedly, without ever rolling down the other side. This is shown for a couple of cycles in the snapshots in Fig.~\ref{Fig-7}(b)-\ref{Fig-7}(d), captured in a frame moving with the velocity ($v$) of the paddlewheel.  This eventually can lead to complete reversal of momentum in the rest frame causing total reflection, even when the incident kinetic energy significantly exceeds the potential barrier.

\section{Conclusions and Outlook}

Our study of the paddlewheel has demonstrated that single-mode scattering studies can reveal features that are suppressed in typical multimode fermionic studies of quantum pumps. The wavepacket approach used here has the distinct advantage of treating classical, semiclassical and quantum dynamics \emph{equivalently}, thus allowing for direct comparison. That has been crucial in our ability to unambiguously identify our anomalous scattering effects, as being essentially classical. That, for example, distinguishes it from non-classical counter-stirring effects produced by a piston like motion of a barrier in a closed ring, dubbed quantum-stirring \cite{Sela-Cohen-PRB-2008}. In this context, it is also important to note there is nothing unusual in itself about net current flow in opposition to the motion of the barrier as can be seen from Fig.~2(b) for a snowplow pump; it simply depends on whether reflection or transmission dominates \cite{das-opatrny}. What is unusual about the paddlewheel pump, is \emph{not} the direction of the net current, but the appearance of prominent scattering peaks at anomalous momenta that can sometimes be the complete opposite of what one expects from a simpler idealized analysis, such as assuming delta-function barriers as often done in theoretical studies of quantum pumps.

The paddlewheel quantum pump as described here can be directly implemented with available technology in ultracold atoms. In fact, we showed a few years ago \cite{Das-Aubin-PRL2009} that mesocopic quantum pumps can be emulated with fermionic cold atoms in narrow waveguides connecting two reservoirs. Recently, using a somewhat different approach, mesoscopic conduction  was indeed demonstrated experimentally \cite{Essilinger} with ultracold fermionc atoms. It would be actually simpler to use wavepackets of BEC instead as assumed here, since there would be no need for reservoirs at the end of the channels as when one literally mimics mesoscopic transport. There has already been progress in implementing the wavepacket approach in experiments on scattering by time-varying potentials, as described in a recent paper \cite{Das-single-barrier} some of us were involved in. That paper focused on an oscillating barrier, but the same setup can be used to create a paddlewheel pump, by introducing periodic translation of the barrier.  While a variety of cold atoms can be used, we assume the same atom species $^{39}K$ atoms in the $|F=1, m_F=+1\rangle$ state, being used in these experiments.  Assuming the transverse harmonic trap frequency of the waveguide to be $\omega_r=2\pi\times 600$ Hz, our units acquire values of $\epsilon=\hbar\omega_r=29\ $nK, $l=\sqrt{\hbar/m\omega_r}=0.65 \mu$m and $\tau=\omega_r^{-1}=0.26$ ms. So our chosen physical parameters become: barriers width $\sigma=10 l=6.5 \mu$m, barrier amplitude $U_0=1\epsilon=29$ nK, oscillation frequency $\omega_{osc}=0.2\omega_r=2\pi\times 120$ Hz and velocity $v=1.5 l/\tau=3.8 $ mm/s. These parameters are accessible in experiments (compare with Table I in Ref. \cite{Das-single-barrier}). The quantum regime can be explored with wavepackets of BEC with interactions suppressed by Feshbach resonances, available for most species for which BEC has been created, including $^{39}K$ \cite{Grimm-RMP-Feshbach}. The classical regime can use non-degenerate wavepackets.

Considering the dearth of experiments on quantum pumps, implementing a paddlewheel pump with cold atoms will provide a setup for studying many properties of quantum pumps that have only been studied theoretically so far. The anomalous scattering effects described here demonstrate that a paddlewheel pump can motivate the general study of scattering of BEC's by time-varying potentials, something that has not been explored yet. The paddlewheel pump  can find use in coherent atomtronics \cite{Das-wavepacket,Holland-PRL-atomtronics} transport, and concrete data gained from experiments can provide insights useful for implementing electronic quantum pumps.  The rich scattering dynamics reveals intriguing interplay between quantum and classical physics and as such there is much potential for using this to probe the quantum versus classical paradigms. Variations of the paddlewheel pump can lead to interesting physics beyond what is described here including: using a well instead of a barrier which can lead to long term trapping of particles and yield chaotic trajectories; exploring the effects of nonlinearity possible with BECs; optimization of selective transport of particles or energy; application to spin pumping.
\vfill

\section{Acknowledgments} We acknowledge valuable discussions with S. Aubin, T. Byrd and J. Delos, and support from the National Science Foundation under Grants No. PHY-0970012, No. PHY-1313871 and No. PHY11-25915.


\end{document}